\documentclass[conference]{IEEEtran}
\IEEEoverridecommandlockouts

\usepackage{cite}
\usepackage{amsmath,amssymb,amsfonts}
\usepackage{algorithmic}
\usepackage{algorithm}
\usepackage{graphicx}
\usepackage{textcomp}
\usepackage{xcolor}
\usepackage{cancel}
\usepackage{colonequals}
\usepackage[caption=false,font=normalsize,labelfont=sf,textfont=sf]{subfig}
\def\BibTeX{{\rm B\kern-.05em{\sc i\kern-.025em b}\kern-.08em
    T\kern-.1667em\lower.7ex\hbox{E}\kern-.125emX}}
\begin{document}

\title{On the Fractional Fourier Transform for\\FMCW Radar Interference Mitigation
\thanks{Research funded by the Austrian Research Promotion Agency (FFG), Infineon Technologies Austria AG and Graz University of Technology under the REPAIR project (40352729)}
\thanks{© 2025 IEEE. Personal use of this material is permitted. Permission from IEEE must be obtained for all other uses, in any current or future media, including reprinting/republishing this material for advertising or promotional purposes, creating new collective works, for resale or redistribution to servers or lists, or reuse of any copyrighted component of this work in other works.}
\thanks{This work has been published at 2025 IEEE Radar Conference (RadarConf25). DOI: 10.1109/RadarConf2559087.2025.11205018}}

\author{\IEEEauthorblockN{Christian Oswald}
\IEEEauthorblockA{\textit{Graz University of Technology} \\
Graz, Austria \\
https://orcid.org/0009-0006-9898-5865}
\and
\IEEEauthorblockN{Josef Kulmer}
\IEEEauthorblockA{\textit{Infineon Technologies Austria AG} \\
Graz, Austria \\
https://orcid.org/0000-0001-9667-3135}
\and
\IEEEauthorblockN{Franz Pernkopf}
\IEEEauthorblockA{\textit{Graz University of Technology} \\
Graz, Austria \\
https://orcid.org/0000-0002-6356-3367}
}

\maketitle

\begin{abstract}
In this paper, we extend our method \cite{oswald2026fmcw} for FMCW radar mutual interference mitigation (IM) based on the discrete fractional Fourier transform (DFrFT). Firstly, we propose a radar signal processing chain including our DFrFT-based IM for real-valued receivers, which we compare to reference algorithms on a synthetic data set. We then reduce computational complexity by reformulating DFrFT-based IM in terms of sparse update signals, which enables mitigation of multiple interferences simultaneously. Finally, we conduct a case study on measurement data and show that our method is compatible with real-world environments.
\end{abstract}


\section{Introduction}

In the automotive industry, FMCW radar is an essential component of advanced driver assistance systems, enabling applications such as adaptive cruise control, forward collision or lane departure warning systems. However, multiple such sensors might interfere with each other when transmitting in the same frequency band. The sensor signal corruptions caused by mutual interference are manifold and range from increased noise floors to ghost objects \cite{toth2018analytical}. 
This necessitates the development of mutual interference mitigation schemes.

\begin{figure}[t]
\centering
\includegraphics[width=0.9\columnwidth]{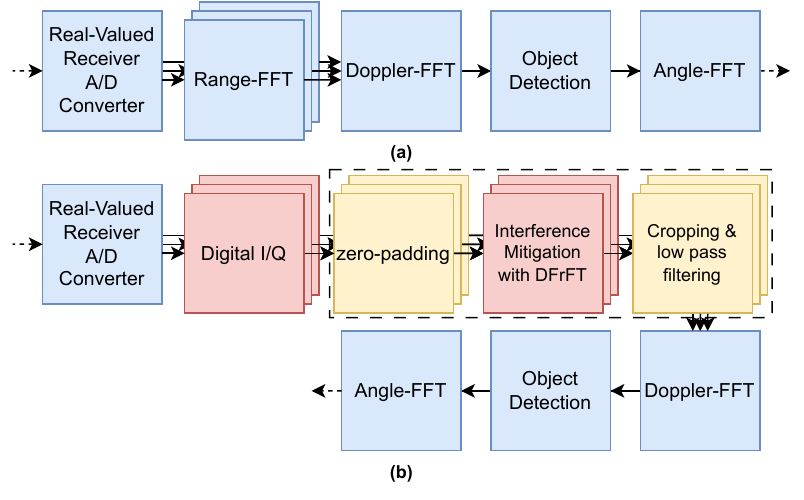}
\caption{FMCW radar signal processing chain for real-valued receivers without \textbf{(a)} and with \textbf{(b)} our proposed IM scheme. Our algorithm outputs range-spectra, meaning that we can remove the range-FFTs. The yellow blocks depict additional processing steps for our padding scheme described in Sec. \ref{sec:main_method} and \cite{oswald2026fmcw}. If our method is used without padding, the yellow blocks are omitted. The dashed box encompasses all processing steps which are included in Alg. \ref{alg1}.}
\label{fig:proc_chain}
\vspace{-\baselineskip}
\end{figure}

Most radar architectures contain in-phase and quadrature (I/Q) receivers \cite{richards2005fundamentals} for reliable detection of echoes. However, the large-scale deployment of automotive radar sensors has driven the development of real-valued receivers. Their production is much cheaper because they only need one receiver processing chain and circumvent common imperfections such as I/Q-imbalance \cite{richards2005fundamentals}. Therefore, mutual interference mitigation algorithms should be compatible with such radar architectures when designed for the automotive industry. The standard signal processing chain for FMCW radar with real-valued receivers is shown in Fig. \ref{fig:proc_chain}a.

Many countermeasures against FMCW radar mutual interference have already been proposed. Besides interference avoidance \cite{bechter2016bats}, possible IM schemes include zeroing interferences in the time-domain \cite{fischer2016untersuchungen}, nonlinear filtering of range-spectra \cite{wagner2018threshold}, variational signal separation \cite{toth2024variational}, adaptive noise cancellation \cite{jin2019automotive} and data driven approaches \cite{ristea2020fully, oswald2023angle, oswald2023end}, among others. A more comprehensive analysis of existing IM algorithms can be found in \cite{toth2019performance}.

The main contributions of our paper are: (i) We review key background concepts necessary to understand our approach (Sec. \ref{sec:background}). (ii) We propose a novel processing chain for real-valued receivers that incorporates DFrFT-based IM (Sec. \ref{sec:main_method}). (iii) We enhance the computational efficiency of the proposed method (Sec. \ref{sec:eigenspace_alg}). (iv) We validate our approach through experiments on both synthetic real-valued data and measured I/Q data (Sec. \ref{sec:experiments}).

We denote matrices with boldface uppercase and vectors as well as sets with boldface lowercase letters. $\boldsymbol{A}[n,m]$ references the element of $\boldsymbol{A}$ in row $n$ and column $m$. $\boldsymbol{A}[:,m]$ is a column vector consisting of the $m^\textrm{th}$ column of $\boldsymbol{A}$. $\boldsymbol{A}[n]$ is a column vector constructed from the $n^\textrm{th}$ row of $\boldsymbol{A}$. Our notation and symbols used are consistent with \cite{oswald2026fmcw}.

\section{Background} \label{sec:background}
\subsection{Signal Model} 
We model radar signals $\boldsymbol{s}$ of length $N$ as a superposition of an object signal $\boldsymbol{s}_O$ with $N_I$ interferences $\boldsymbol{s}_I$ and noise $\boldsymbol{s}_\mathcal{N}$,
\begin{equation}
    \boldsymbol{s} = \boldsymbol{s}_O + \sum_{m=1}^{N_I}  \boldsymbol{s}_{I_m} + \boldsymbol{s}_\mathcal{N}. \label{eq:superposition}
\end{equation}
Our model assumes I/Q signals, where all terms in \eqref{eq:superposition} are complex-valued; the corresponding signal from a real-valued receiver then simply is $\Re \{\boldsymbol{s}\}$.
An object signal $\boldsymbol{s}_O$ sampled in intervals $T_s$ consists of a superposition of $N_O$ objects,
\begin{equation}
    \boldsymbol{s}_O[n] = \sum_{i=1}^{N_O} A_i e^{j(\omega_i nT_s+\phi_i)},
\end{equation}
where each object is parameterized by its echo amplitude $A_i$, frequency $\omega_i$ and initial phase $\phi_i$.
Mutual interference appears as a complex-valued linearly frequency modulated (LFM) chirp $\boldsymbol{s}_I$ with amplitude $A$ and initial phase $\phi_0$,
\begin{equation}
\begin{split}    
\boldsymbol{s}_I[n] &=
\begin{cases}
A e^{j(-2\pi k\tau n T_s + \pi k n^2 T_s^2+ \phi_0)} & \frac{\tau-B/k}{T_s} < n <\frac{\tau+B/k}{T_s},\\
{0,}&{\text{otherwise,}} 
\end{cases}\label{eq:int_model}
\end{split}
\end{equation}
where $B$ is the bandwidth of the ideal anti-aliasing filter, $k$ the chirp rate of the interference and $\tau$ the time delay when the interferer and the victim radar's instantaneous transmit frequency cross.
We collect all sources of noise and clutter in $\boldsymbol{s}_\mathcal{N}$ modelled as additive complex-valued white Gaussian noise. Our signal model is identical to \cite{oswald2026fmcw}.
\subsection{The Discrete Fractional Fourier Transform}
The discrete fractional Fourier transform (DFrFT) of fractional order $2 \alpha/\pi, \alpha \in \mathbb{R},$ is a generalization of the the discrete Fourier transform (DFT) $\boldsymbol{W}$, which can constructed via its eigendecomposition
\begin{equation}
    \boldsymbol{W}^{\frac{2\alpha}{\pi}} = \boldsymbol{V}\boldsymbol{\Lambda}^{\frac{2\alpha}{\pi}}\boldsymbol{V}^T, \label{eq:eigendecomp}
\end{equation}
where $\boldsymbol{V}$ and $\boldsymbol{\Lambda}$ are the DFT eigenvectors and eigenvalues, respectively. For more concise notation, we write $\boldsymbol{W}_\alpha \colonequals \boldsymbol{W}^{\frac{2\alpha}{\pi}}$ and $\boldsymbol{\Lambda}_\alpha \colonequals \boldsymbol{\Lambda}^{\frac{2\alpha}{\pi}}$, where $\alpha$ is the so-called fractional angle. The DFT eigenvectors are not unique, but if $\boldsymbol{V}$ is chosen to approximate the Hermite-Gauss functions, $\boldsymbol{W}_\alpha$ compresses a complex-valued LFM chirp with chirp rate $\cot (\alpha)$ into a small number of samples. Equivalently, such a DFrFT can be interpreted as rotating a signal's time-frequency representation by $\alpha$ radians; an example can be seen in Fig. \ref{fig:stfts_madfrft}a and Fig. \ref{fig:stfts_madfrft}b. Surveys on the DFrFT and its applications can be found in \cite{su2019analysis, gomez2020fractional}.

\subsection{Interference Mitigation using the DFrFT} \label{sec:old_method}
In \cite{oswald2026fmcw} we proposed a method that compresses, detects and subsequently zeroes LFM interference chirps in I/Q signals using the DFrFT. We simultaneously compute the DFrFTs of $M$, $N \textrm{ mod } M = 0$, fractional angles equally spaced between $- \pi$ and $\pi$ radians with our efficient multiangle DFrFT (EMDFrFT), which is an adaptation of the multiangle centered DFrFT in \cite{vargas2005multiangle}. The EMDFrFT $\boldsymbol{S}$ of a signal $\tilde{\boldsymbol{s}}$ is given as
\begin{gather}
\boldsymbol{\rho} = \boldsymbol{V}^T \tilde{\boldsymbol{s}}, \label{eq:rho} \\
\bar{\boldsymbol{Z}}[p,n] = \boldsymbol{V}^T[p,n] \boldsymbol{\rho} [p], \label{eq:z} \\
\boldsymbol{S}[m,n] = \textrm{FFT}_m \left\{\sum_{l=0}^{\frac{N}{M}-1} \bar{\boldsymbol{Z}}[m + lM, n]\right\}, \label{eq:S}
\end{gather}
where $n,p \in \{0,1,...,N-1\}$, $m \in \{0,1,...,M-1\}$, and $\textrm{FFT}_m\{\cdot\}$ computes column-wise FFTs of its input matrix. Note that the algorithm by \cite{vargas2005multiangle} efficiently computes the multiangle \textit{centered} DFrFT; however, in \cite{oswald2026on} we derive the multiangle (standard) DFrFT with identical complexity. 

We then search for the global maximum $\boldsymbol{S}[\hat{m},\hat{n}]$ within our search space,
\begin{align}
\hat{m}, \hat{n} &= \textrm{arg\,max} \{ \boldsymbol{M} \odot|\boldsymbol{S}[m,n]|\}, \label{eq:argmax}
\end{align}
where $\odot$ denotes element-wise multiplication and $\boldsymbol{M}$ is a binary mask restricting the fractional angles in our search space such that they are bounded by $\pm \alpha_\text{max}$, which is hyperparameter of our method. We then classify $\boldsymbol{S}[\hat m, \hat n]$ using a least-of constant false alarm rate (LO-CFAR) detector. This detector uses $\Phi$ samples of $\boldsymbol{S}[\hat m]$ surrounding $\boldsymbol{S}[\hat{m}, \hat{n}]$ with a distance of $G$ guard cells as its noise estimation window. If the signal-to-noise ratio (SNR) corresponding to $\boldsymbol{S}[\hat{m}, \hat{n}]$ is above the detector's threshold $\beta$, we set $\boldsymbol{S}[\hat{m}, \hat{n}]$ and the surrounding $G$ guard cells to zero; we formalize this zeroing operation with a binary detection mask $\boldsymbol{d}$,
\begin{align}
\boldsymbol{d} &= \textrm{LO-CFAR}(\boldsymbol{S}[\hat m],\hat n), \label{eq:CFAR}
\end{align}
where $\boldsymbol{d}[n]$ is 0 for $n$ we want to zero and 1 otherwise, and restart the process at \eqref{eq:rho} with an updated \mbox{$\tilde{\boldsymbol{s}} \leftarrow \boldsymbol{d} \odot \boldsymbol{S}[\hat{m}]$}. We loop this procedure of compression, detection and zeroing until the SNR corresponding to the global maximum is below $\beta$; then we terminate the algorithm and return the interference mitigated range-spectrum.
We expect this algorithm to loop $N_I$-times on a fast-time sequence $\boldsymbol{s}$ contaminated by $N_I$ interference chirps. Assuming $M \ll N$, the algorithm's computational complexity is $\mathcal{O}(N_IN^2/2)$, since the most expensive operation is the matrix-multiplication \eqref{eq:rho}; as shown in \cite{vargas2005multiangle}, $\boldsymbol{V}$ is symmetric, which allows us to roughly halve the complexity from a dense matrix multiplication to $\mathcal{O}(N^2/2)$ -- the same optimization is possible for \eqref{eq:z}. Meanwhile, we showed in \cite{oswald2026on} that we only require approximately $N/2$ instead of $N$ FFTs to compute $\eqref{eq:S}$. A detailed explanation of our method as well as pseudo-code can be found in \cite{oswald2026fmcw}.

\section{Real-Valued Interference Mitigation} \label{sec:main_method}

\begin{figure*}[!t]
\centering
\includegraphics[width=\textwidth]{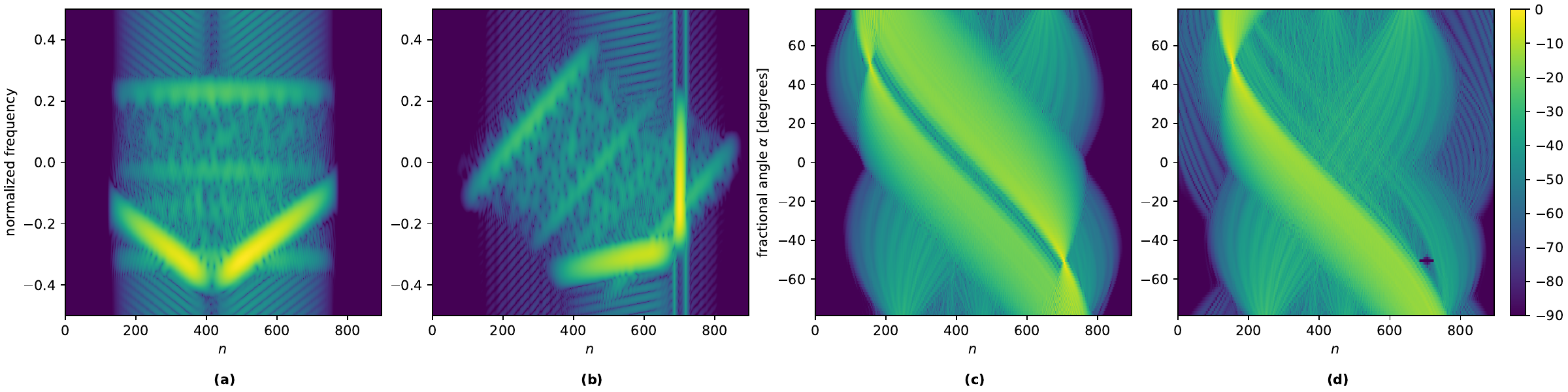}

\caption{\textbf{(a)} STFT of an interfered digital I/Q signal $\boldsymbol{s}$ and \textbf{(b)} STFT of $\boldsymbol{W}_{\hat{\alpha}} \boldsymbol{s}$ with $\hat\alpha \approx -50^\circ$ rotating the time-frequency representation of $\boldsymbol{s}$. $\boldsymbol{W}_{\hat{\alpha}} \boldsymbol{s}$ compresses one of the LFM chirps into a pulse, which is also visible at the corresponding location in \textbf{(c)}, the EMDFrFT magnitudes $|\boldsymbol{S}|$ of $\boldsymbol{s}$. If we now zero that compressed chirp and recompute the EMDFrFT, we get \textbf{(d)}; notice how the peak corresponding to the other chirp has not changed, as the two interference chirps are \textit{separable}. Therefore, we can zero both chirps in the same iteration of our algorithm as explained in Sec. \ref{sec:parallel}; $\boldsymbol{s}$ has been padded according to Sec. \ref{sec:main_method}. All plots are in dB and renormalized such that the maximum value is 0.}
\label{fig:stfts_madfrft}
\end{figure*}

Since every real-valued signal is equivalent to the sum of a complex-valued signal and its complex conjugate, the so-called \textit{image component}, the EMDFrFT of a real-valued signal is Hermite-symmetric about a fractional angle of 0 and 180 degrees. In other words, the DFrFT cannot compress real-valued LFM chirps into pulses because it has complex-valued LFM chirps as its basis functions.
This fundamental problem is closely related to the reduced SNR in the output of matched filters applied to real-valued or I/Q-imbalanced signals, where the image component appears as an increased noise floor \cite{richards2005fundamentals, sinsky1974error}. 
Furthermore, a complex-valued interference chirp and its image component cannot be separated at any fractional angle, which means that zeroing one component alters the appearance of its image. The image therefore becomes an \textit{incomplete interference} as discussed in \cite{oswald2026fmcw}, which renders it undetectable in the worst case. Therefore, we propose to use digital I/Q demodulation \cite{richards2005fundamentals} as a preprocessing step before DFrFT-based interference mitigation. Our proposed processing chain is depicted in Fig. \ref{fig:proc_chain}b.

Digital I/Q demodulation generates I and Q components from a real-valued receive signal \textit{after} A/D conversion; more specifically, it discards one sideband and centers its complementary half around DC. There exist various architectures which efficiently implement such a scheme \cite{rader1984simple, shaw1995q, richards2005fundamentals}. After digital I/Q, real-valued interference chirps appear as two consecutive complex-valued chirps, which we can mitigate independently. An interfered signal after digital I/Q and its EMDFrFT are depicted in Fig. \ref{fig:stfts_madfrft}a and Fig. \ref{fig:stfts_madfrft}c, respectively. Hypothetically, the real-valued interference chirp's components could also be separated by computing the receive signal's analytic representation via a Hilbert-transform; however, these interference components do not span the entire bandwidth, i.e., they are again incomplete, meaning that they cannot be optimally compressed by any DFrFT. Therefore, digital I/Q is preferable to analytic signals as preprocessing for DFrFT-based IM. Since interferences mostly appear as V-shapes in the short-time Fourier transform (STFT) of the digital I/Q signal, we expect our DFrFT-based IM method to iterate, at maximum, twice as often as on the corresponding I/Q received data. We deal with this increased computational load in Sec. \ref{sec:eigenspace_alg} by reformulating our algorithm in terms of sparse update signals, enabling simultaneous processing of these \textit{separable} chirps such that the number of iterations is the same as on the corresponding I/Q received signal.

In \cite{oswald2026fmcw} we observed that signal components located at the corners of its time-frequency representation cannot be properly compressed by a DFrFT; therefore, we proposed increasing the A/D-converter's sampling rate and zero-padding the input signal before computing its EMDFrFT. Oversampling can be implemented by adapting processing blocks within architectures such as \cite{rader1984simple, shaw1995q}.
Such oversampling avoids that digital I/Q maps DC components of the real-valued receive signal to the Nyquist frequency. 
An example for a sufficiently padded digital I/Q signal can be seen in Fig. \ref{fig:stfts_madfrft}a. 

\section{Increased Efficiency through Mitigation in the DFT Eigenbasis} \label{sec:eigenspace_alg}
In this section we reformulate our DFrFT-based IM algorithm \cite{oswald2026fmcw} to mitigate \textit{separable} interferences simultaneously while computing successive iterations more efficiently. 
As derived in \cite{oswald2026fmcw}, the time-domain signals $\boldsymbol{s}_{i}$ and $\boldsymbol{s}_{i+1}$ before iterations $i$ and $i+1$ of our algorithm are related by
\begin{equation}
\begin{split}
    \boldsymbol{s}_{i+1} &=\boldsymbol{W}_{-\hat\alpha_i} (\boldsymbol{d}_i \odot (\boldsymbol{W}_{\hat\alpha_i} \boldsymbol{s}_i)) \\
    &= \boldsymbol{s}_i - \boldsymbol{W}_{-\hat\alpha_i} ((1 - \boldsymbol{d}_i) \odot (\boldsymbol{W}_{\hat\alpha_i} \boldsymbol{s}_i)),
\end{split}
\end{equation}
where the binary mask $\boldsymbol{d}_i$ zeroes some compressed interference chirp with chirp rate $\cot (\hat\alpha_i)$. We define \mbox{$\boldsymbol{\gamma}_i = (1 - \boldsymbol{d}_i) \odot (\boldsymbol{W}_{\hat\alpha_i} \boldsymbol{s}_i)$} for more concise notation; the eigen-coefficients $\boldsymbol{\rho}_i = \boldsymbol{V}^T\boldsymbol{s}_{i}$ and $\boldsymbol{\rho}_{i+1} = \boldsymbol{V}^T\boldsymbol{s}_{i+1}$ relate via
\begin{equation}  
\begin{split}
    \boldsymbol{\rho}_{i+1} &= \boldsymbol{\rho}_i - \boldsymbol{V}^T\boldsymbol{W}_{-\hat\alpha_i} \boldsymbol{\gamma}_i \\
     &\overset{\eqref{eq:eigendecomp}}{=} \boldsymbol{\rho}_i - \boldsymbol{V}^T\boldsymbol{V} \boldsymbol{\Lambda}_{-\hat\alpha_i} \boldsymbol{V}^T \boldsymbol{\gamma}_i \\
     &= \boldsymbol{\rho}_i - \boldsymbol{\Lambda}_{-\hat\alpha_i} \boldsymbol{V}^T \boldsymbol{\gamma}_i, \label{eq:projections}
\end{split}
\end{equation}
where $\boldsymbol{V}^T\boldsymbol{V}$ is the identity due to $\boldsymbol{V}$ being orthonormal \cite{oswald2026fmcw}. This means that we can compute iterations of our algorithm by subtracting a scaled projection of the detected interference chirp from $\boldsymbol{\rho}_i$, which has already been evaluated in the previous iteration. This is more efficient than the formulation presented in \cite{oswald2026fmcw}, because \mbox{$\boldsymbol{V}^T \boldsymbol{\gamma}_i$} has complexity $\mathcal{O}((2G+1) \cdot N)$, $N \gg G$, as $1 -\boldsymbol{d}_i$ sets all except $2G+1$ values of $\boldsymbol{W}_{\hat\alpha_i} \boldsymbol{s}_i$ to zero; 
meanwhile, \cite{oswald2026fmcw} evaluates $\boldsymbol{V}^T (\boldsymbol{d}_i \odot (\boldsymbol{W}_{\hat\alpha_i} \boldsymbol{s}_i))$ in every iteration which has complexity $\mathcal{O}((N-2G-1) \cdot N)$ (when ignoring the special structure of $\boldsymbol{V}$). In our experiments conducted in Sec. \ref{sec:experiments}, $G=10$ while $N=896$, which means that our new formulation results in a $42$-fold reduction in the number of operations for computing $\boldsymbol{\rho}_{i+1}$ when $i \geq 1$. 
For $i = 0$, we still need to compute $\boldsymbol{\rho}_1 = \boldsymbol{V}^T \boldsymbol{s}_1$ with complexity $\mathcal{O}(N^2/2)$. 
Since $\boldsymbol{\Lambda}_{-\hat\alpha_i}$ is a diagonal matrix, its product with \mbox{$\boldsymbol{V}^T \boldsymbol{\gamma}_i$} has negligible complexity. Overall, the computational complexity remains $\mathcal{O}(N_IN^2/2)$ because of \eqref{eq:z}, which consists of $N^2/2$ multiplications \cite{vargas2005multiangle} and is now the most expensive operation in our algorithm. 

Since all $\boldsymbol{\rho}$ are projections of time-domain signals, the circular shifting of the row indices in successive iterations as described in \cite{oswald2026fmcw} vanishes.
This is because in \eqref{eq:projections} the inverse DFrFT matrix $\boldsymbol{W}_{-\hat{\alpha}_i}$ transforming $\boldsymbol{\gamma}_i$ back into the time-domain becomes the diagonal matrix $\boldsymbol{\Lambda}_{-\hat\alpha_i}$ in the DFT eigenbasis. 
\subsection{Mitigating multiple interferences simultaneously} \label{sec:parallel}
As discussed in Sec. \ref{sec:main_method}, after digital I/Q, real-valued LFM chirp interference appears as two consecutive complex chirps which we can mitigate independently. In other words, there exists a fractional Fourier domain where these two chirps are \textit{separable} since they do not cross in the time-frequency plane. 
For instance, if we zero out one interference component in Fig. \ref{fig:stfts_madfrft}b according to our method and then compute another EMDFrFT depicted in Fig. \ref{fig:stfts_madfrft}d, we notice that the appearance of the other interference component has not changed. Therefore, we can simply zero both interference components in the same iteration. In fact, such separable interferences also appear in I/Q received data; examples would be a signal interfered by two chirps with the same chirp rate, or the interferences in Fig. 2c of \cite{oswald2026fmcw}. We implement simultaneous mitigation by looping \eqref{eq:argmax} and \eqref{eq:CFAR} until the detector returns \textit{no interference}. After every iteration of this new inner loop we mask $\boldsymbol{S} \leftarrow \boldsymbol{M} \odot \boldsymbol{S}$ to only include chirps $\boldsymbol{S}[m,n]$ that are separable from the previous global maximum\footnote{A possible variation of this procedure would be to first run the CFAR-detector on all entries of $\boldsymbol{S}$ and then loop \eqref{eq:argmax} followed by masking on all SNRs that are above the detection threshold.}. $\boldsymbol{M}$ is constructed from a global maximum $\boldsymbol{S}[\hat m, \hat n]$ by setting all $\boldsymbol{M}[m,n]$ to zero where the chirp corresponding to $\boldsymbol{S}[\hat m, \hat n]$ is present, while all other entries are one; we denote this mask construction procedure as NotSupportOfChirp$()$ in Alg. \ref{alg1}. 
All possible $\boldsymbol{M}$ corresponding to all possible pairs of $\hat m$ and $\hat n$ can be precomputed and stored in lookup tables; we only need to store $N/2$ such masks (where masks corresponding to neighboring column indices are almost identical) thanks to the rotation symmetry within the EMDFrFT. 
We collect all $N_p$ separable interferences in their respective fractional domains $\boldsymbol{\hat\alpha}_i$ in $\boldsymbol{\Gamma}_i$.
All chirps are then mitigated simultaneously through

\begin{equation}
    \boldsymbol{\rho}_{i+1} = \boldsymbol{\rho}_i - \sum_{k=0}^{N_p -1} \boldsymbol{\Lambda}_{-\hat{\boldsymbol{\alpha}}_i[k]} \boldsymbol{V}^T \boldsymbol{\Gamma}_i[k] 
    \label{eq:parallel_update}
\end{equation} 
replacing \eqref{eq:projections}.
Note that this parallelism is enabled by \eqref{eq:projections} and is not possible with the formulation in \cite{oswald2026fmcw}. 
Mitigating separable interferences simultaneously using \eqref{eq:parallel_update} is equivalent to mitigating one interference per iteration using \eqref{eq:projections}. However, the LO-CFAR detector might return false negatives due to the increased noise estimate caused by the other interferences; in that case, the interference will only be detected in the next iteration. The false negative rate could be lowered by designing more elaborate detectors. Further optimizations are possible for digital I/Q signals where, due to their V-shape, the two peaks corresponding to the two LFM chirps always appear as predictable pairs within the EMDFrFT;  we leave these ideas for further research. Our method's overall computational complexity is $\mathcal{O}(N^2/2)$ if all $N_I$ interferences are separable and $\mathcal{O}(N_IN^2/2)$ if none of them are separable. 
Note that this improved formulation can still be combined with other potential speed-ups listed in \cite{oswald2026fmcw}, for example by using DFT eigenvectors with sparse and repeating entries \cite{de2019reduced} or eigenvectors such as \cite{erseghe2003orthonormal} where the change-of-basis can be computed in $\mathcal{O}(N \log N)$ \cite{erseghe2006efficient}.

\subsection{Algorithm}
Our method including simultaneous processing of separable interferences is summarized in Alg. \ref{alg1}. The $M \times N$ matrix $\boldsymbol{K}$ implements the sum within \eqref{eq:S} and consists of $N/M$ concatenated $M \times M$ identity matrices. We stack $\boldsymbol{\rho}$ $N/2-$times resulting in the $N \times N/2$ matrix $[\boldsymbol{\rho} \; \ldots \;\boldsymbol{\rho}]$ to implement \eqref{eq:z} with optimizations from \cite{vargas2005multiangle}, while Rearrange$()$ indicates the reconstruction of the full $M \times N$ EMDFrFT after computing $N/2$ $M$-point FFTs as described in \cite{oswald2026on}. We apply the window function $\boldsymbol{w}$ to every $\boldsymbol{s}$ before DFrFT-based IM.
All hyperparameters are listed in Tab. \ref{tab:dfrft}. 

\begin{algorithm}[H]
\caption{Simultaneous IM in the DFT Eigenbasis}
\begin{algorithmic}
\STATE {\textsc{IMfracV2}}$(\boldsymbol{s})$: \COMMENT{$\boldsymbol{s}$ is possibly interfered}
\STATE $\text{initialize } \boldsymbol{\alpha}$ \COMMENT{array of the fractional angles evaluated} 
\STATE $\text{initialize } \boldsymbol{M}$ \COMMENT{mask for restricting $\boldsymbol{S}$, see \cite{oswald2026fmcw} and Sec. \ref{sec:parallel}}
\STATE $\text{initialize } m_{RS}$ \COMMENT{row index of range-spectrum in $\boldsymbol{S}$}
\STATE $\boldsymbol{s} \gets \boldsymbol{s} \odot \boldsymbol{w}$ \COMMENT{apply window function}
\STATE $\boldsymbol{s} \gets \mathrm{ZeroPad(\boldsymbol{s})}$ \COMMENT{optional, see \cite{oswald2026fmcw} and Sec. \ref{sec:main_method}}
\STATE $\boldsymbol{\rho} = \boldsymbol{V}^T \boldsymbol{s}$ \COMMENT{initial change-of-basis}
\STATE $ \textbf{do} $
\STATE \hspace{0.5cm}$\boldsymbol{S} \gets \textrm{Rearrange}(\textrm{FFT}_m\{ \boldsymbol{K} (\boldsymbol{V}^T \odot  [\boldsymbol{\rho} \; \ldots \;\boldsymbol{\rho}])\})$
\STATE \hspace{0.5cm}$ \textbf{do } $ \COMMENT{All updates for $\boldsymbol{\rho}$ can be computed in parallel}
\STATE \hspace{1cm}$ \hat m, \hat n  \gets \textrm{arg max}|\boldsymbol{M} \odot \boldsymbol{S}|$
\STATE \hspace{1cm}$ \boldsymbol{d} \gets \textrm{LO-CFAR}(\boldsymbol{S}[\hat m],\hat n)$ \COMMENT{$\boldsymbol{d}$ is a binary mask}  
\STATE \hspace{1cm}$ \boldsymbol{\rho} \gets \boldsymbol{\rho} - \boldsymbol{\Lambda}_{-\boldsymbol{\alpha}[\hat m]} \boldsymbol{V}^T ((1 - \boldsymbol{d}) \odot \boldsymbol{S}[\hat m])$
\STATE \hspace{1cm}$ \boldsymbol{M} \gets \boldsymbol{M} \land \textrm{NotSupportOfChirp}(\hat m, \hat n)$
\STATE \hspace{0.5cm}$ \textbf{while } \boldsymbol{d}\textrm{ contains a detection (i.e., a zero)} $
\STATE \hspace{0.5cm}$ \boldsymbol{M} \gets \textrm{reset}(\boldsymbol{M})$ \COMMENT{restrict $\boldsymbol{S}$ to angles $\boldsymbol{\alpha}$, see \cite{oswald2026fmcw}}
\STATE $ \textbf{while } \boldsymbol{\rho}\textrm{ has changed in previous iteration} $
\STATE $ \boldsymbol{s}_{RS} \gets \textrm{Crop}(\boldsymbol{S}[m_{RS}]) $ \COMMENT{optional, see \cite{oswald2026fmcw} and Sec. \ref{sec:main_method}}
\STATE $ \boldsymbol{s}_{RS} \gets \textrm{LowPass}(\boldsymbol{s}_{RS}) $ \COMMENT{optional, see \cite{oswald2026fmcw} and Sec. \ref{sec:main_method}}
\STATE \textbf{return} $\boldsymbol{s}_{RS}$ \COMMENT{interference mitigated range-spectrum}
\end{algorithmic}
\label{alg1}
\end{algorithm}

\section{Experiments} \label{sec:experiments}

\subsection{Performance on synthetic digital I/Q dataset} \label{sec:synthetic}
We evaluate our proposed processing chain including IM for real-valued receivers, labelled \textit{IMfrac}, on a synthetic dataset consisting of 250 interfered and ground-truth range-Doppler (RD) maps. The dataset has the same parameters as the dataset in \cite{oswald2026fmcw}, except that we increased the number of fast-time samples from 512 to 1024 and only use their real part. We generated the corresponding digital I/Q signals of length 512 by discarding the lower sideband via an intermediate forward and inverse FFT. Furthermore, we used a steep high-pass filter to simulate the DC suppression of digital I/Q processing chains such as \cite{rader1984simple, shaw1995q} that eliminate the side effects of non-ideal mixers \cite{richards2005fundamentals}. 
Furthermore, we generated padded digital I/Q signals of length $896$ by oversampling and zero-padding. 
Fig. \ref{fig:stfts_madfrft}a is one interfered fast-time sequence from our dataset.
In addition to Alg. \ref{alg1} we evaluate an oracle variant of \textit{IMfrac}, which we describe in \cite{oswald2026fmcw} in more detail. We compare our processing chain to ramp filtering \cite{wagner2018threshold} and zeroing \cite{fischer2016untersuchungen} with oracle as well as envelope change point interference detection and evaluate the mean squared error (MSE), the signal-to-interference-plus-noise (SINR) ratio, the error vector magnitude (EVM), the true positive rate (TPR), the false alarm rate (FAR) and the F1-score per RD map in the dataset.
The hyperparameters of \textit{IMfrac} in Tab. \ref{tab:dfrft}, the reference IM methods and metrics are the same as in \cite{oswald2026fmcw}. 
Compared to \cite{oswald2026fmcw}, zeroing has been altered such that interference detection is performed on the signals' envelope. 

\begin{figure*}[!t]
\centering
\includegraphics[width=\textwidth]{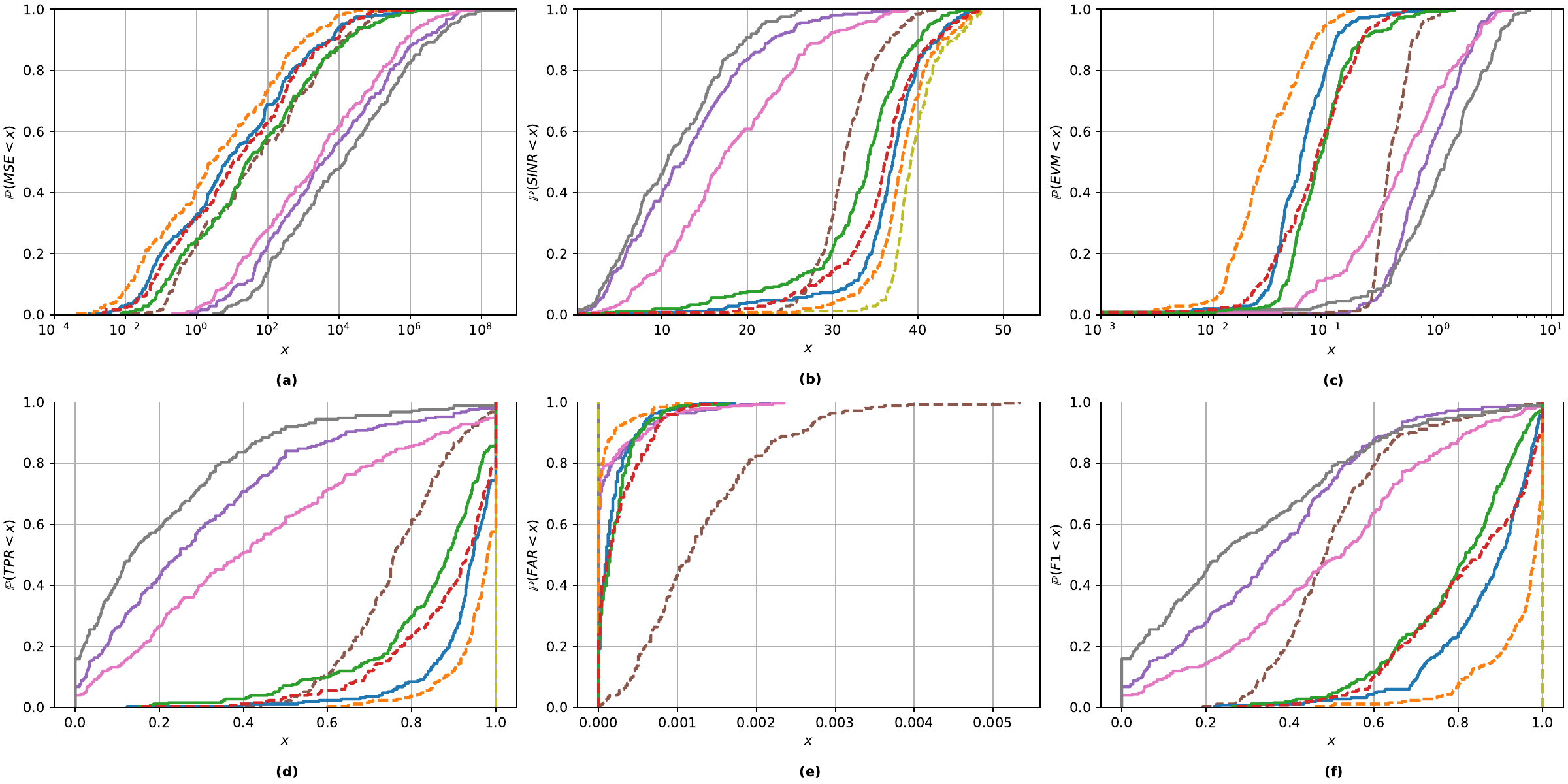} \\
\includegraphics[width=12cm]{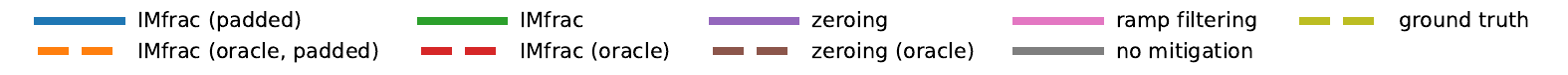}
\caption{Empirical cumulative density functions (ECDFs) of all evaluated metrics per range-Doppler map. The oracle methods are drawn with dashed lines. Note that we have zoomed into relevant parts of the ECDFs to better resolve close-by curves.}
\label{fig:ecdfs}
\end{figure*}

The results are collected in Fig. \ref{fig:ecdfs}. Our DFrFT-based IM algorithm outperforms the reference methods across all metrics. Compared to the results on I/Q received data in \cite{oswald2026fmcw}, padding becomes more important as digital I/Q data contains high-energy components with high negative frequencies (see Fig. \ref{fig:stfts_madfrft}a for instance).
A low MSE does not necessarily lead to good object detection performance, as visible in Fig. \ref{fig:ecdfs}a and Fig. \ref{fig:ecdfs}f for zeroing with oracle detection. 

\begin{table}[t]
\caption{Parameters of IMfrac}
\begin{center}
\begin{tabular}{|c|c|}
\hline
\multicolumn{2}{|c|}{\textbf{Parameter}}  \\
\hline
\# Computed fractional angles by EMDFrFT ($M$) & 256 \\
\hline
Maximum fractional angle ($\alpha_{\textrm{max}}$) & $80^\circ$ \\
\hline
Window-size interference detector ($\Phi$) & $N/2 -G-1$ \\
\hline
\# Guard cells of interference detector ($G$) & 20 \\
\hline
Threshold of interference detector ($\beta$) & 20 dB \\
\hline
\end{tabular}
\label{tab:dfrft}
\end{center}
\vspace{-\baselineskip}
\end{table}

\begin{figure*}[!t]
\centering
\includegraphics[width=.95\textwidth]{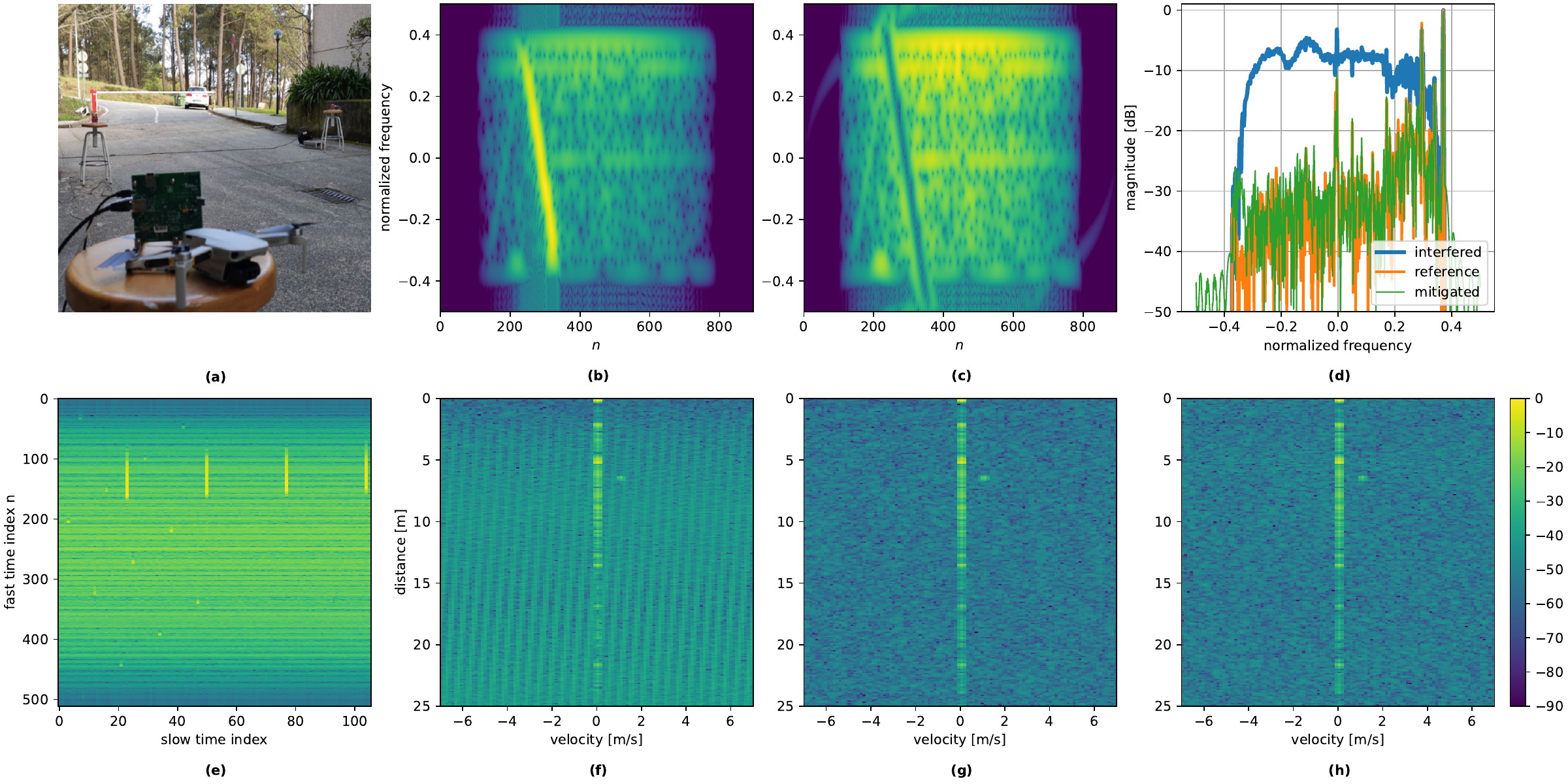}%
\caption{Case study on interfered measurement data from \cite{interference_dataset}. \textbf{(a)} Measurement setup; STFT of ramp \#23 of frame \#2001 before \textbf{(b)} and after \textbf{(c)} interference mitigation with our method. \textbf{(d)} overlay of interfered and interference mitigated range-spectra of frame \#2001 ramp \#23 with reference \#2002 ramp \#23. \textbf{(e)} fast-time/slow-time sequence and \textbf{(f)} RD map of the interfered measurement frame \#2001. \textbf{(g)} RD map of the reference measurement frame \#2002. \textbf{(h)} interference mitigated RD map \#2001. All plots have been normalized such that the maximum value has 0 decibels.}
\label{fig:meas_data}
\vspace{-\baselineskip}
\end{figure*}

\subsection{Case Study on Measurement Data}
In this section we qualitatively analyze our DFrFT-based IM method on I/Q-modulated measurement data provided in \cite{interference_dataset}. We compare measurement frames $\#2001$ and $\#2002$ of the first receiver and assume that the object's locations on the RD maps are the same. 
Since the sensor's number of fast-time samples is $512$, we can apply DFrFT-based IM with the parameter settings from Tab. \ref{tab:dfrft}. We implement our padding scheme by upsampling the fast-time sequences before windowing and zero-padding, which results in $896$ fast-time samples. Our findings are summarized in Fig. \ref{fig:meas_data}.
The measured interference closely resembles an ideal complex-valued LFM chirp, which is easily detected and mitigated. 
Qualitatively, the interference mitigated and the reference RD-maps $\#2001$ and $\#2002$ closely match. 
We hypothesize that the fluctuations in amplitude of the interfered range-spectrum in Fig. \ref{fig:meas_data}h is caused by an imperfect anti-aliasing filter as well as multipath propagation of the interference. 
Although the range-spectra in Fig. \ref{fig:meas_data}h cannot be directly compared, the object peaks previously masked by interference have been reconstructed almost perfectly. 
\section{Conclusion}
In this paper, we have extended our IM method based on the DFrFT \cite{oswald2026fmcw} to real-valued receivers, presented a more efficient formulation and conducted a case study on measurement data. However, we did not yet discuss multipath channels and imperfections of the sensor, such as interferences saturating the receiver. Nevertheless, we are convinced that this paper is an important step towards a practical and performant algorithm for LFM chirp interference mitigation.
\bibliographystyle{IEEEtran}
\bibliography{IEEEabrv,references}

\end{document}